\documentclass[fleqn,usenatbib]{mnras}
\usepackage{multirow}

\usepackage[T1]{fontenc}

\DeclareRobustCommand{\VAN}[3]{#2}
\let\VANthebibliography\thebibliography
\def\thebibliography{\DeclareRobustCommand{\VAN}[3]{##3}\VANthebibliography}

\usepackage{graphicx}	
\usepackage{amsmath}	
\usepackage{amssymb}	

\title[Discovery of Galaxy Clusters with Chandra]{Discovery of Galaxy Clusters and a Head-Tail Radio Galaxy in the Direction of Globular Cluster NGC 6752}

\author[Cheng et al.]{
Zhongqun Cheng,$^{1,2}$\thanks{E-mail: chengzq@whu.edu.cn (WHU)}
Xiaohui Sun,$^{3}$
Wei Wang$^{1,2}$\thanks{E-mail: wangwei2017@whu.edu.cn (WHU)}
and Zhiyuan Li$^{4,5}$
\\
$^{1}$School of Physics and Technology, Wuhan University, Wuhan 430072, People's Republic of China\\
$^{2}$WHU-NAOC Joint Center for Astronomy, Wuhan University, Wuhan 430072, People's Republic of China\\
$^{3}$School of Physics and Astronomy, Yunnan University, 650091 Kunming, People's Republic of China\\
$^{4}$School of Astronomy and Space Science, Nanjing University, Nanjing 210023, People's Republic of China\\
$^{5}$Key Laboratory of Modern Astronomy and Astrophysics (Nanjing University), Ministry of Education, Nanjing 210023,\\ People's Republic of China
}

\date{Accepted XXX. Received YYY; in original form ZZZ}

\pubyear{2015}

\begin{document}
\label{firstpage}
\pagerange{\pageref{firstpage}--\pageref{lastpage}}
\maketitle

\begin{abstract}
We report the discovery of CXOU J191100-595621 and CXOU J191012-595619, two galaxy clusters serendipitously detected in the direction of globular cluster NGC 6752, based on archival {\it Chandra} observations with a total exposure time of $\sim 344$ ks. The deep {\it Chandra} X-ray data enabled us to measure properties of both systems, which result in a redshift of $z=0.239\pm0.013$ and $z=0.375\pm0.016$, a temperature of $kT=3.32^{+0.57}_{-0.46}$ keV and $kT=3.71^{+1.18}_{-0.86}$ keV, an iron abundance of $Z_{\rm Fe}=0.64^{+0.34}_{-0.29}Z_{\rm Fe\odot}$ and $Z_{\rm Fe}=1.29^{+0.97}_{-0.65}Z_{\rm Fe\odot}$, and a rest-frame full band (0.5-7 keV) luminosity of $L_{\rm X}=9.2^{+1.2}_{-1.1}\times 10^{43} {\rm \, erg\, s^{-1}}$ and $L_{\rm X}=9.9^{+2.7}_{-2.2}\times 10^{43} {\rm \, erg\, s^{-1}}$ for CXOU J191100-595621 and CXOU J191012-595619, respectively. The temperature profile of CXOU J191100-595621 is found to decreases with decreasing radius, indicating a cool core in this cluster. The hydrostatic equilibrium estimation suggests the clusters are moderately weighted, with $M_{500}=(1.3\pm0.4)\times 10^{14}\, M_{\odot}$ and $M_{500}=(2.0\pm1.5)\times 10^{14}\, M_{\odot}$, respectively. We search for optical and radio counterparts of X-ray point sources in the clusters. Three active galactic nuclei are found, among which one is identified with a narrow-angle-tail radio galaxy, and one is found to associated with the brightest central galaxy (BCG) of CXOU J191100-595621. 
\end{abstract}




\section{Introduction}
Clusters of galaxies are permeated with the hot and dilute intracluster medium (ICM) emitting thermal bremsstrahlung and metal line emission in X-ray, which make them one of the most luminous X-ray sources ($L_{\rm X}\sim 10^{43-45}\, {\rm erg\, s^{-1}}$) in the Universe \citep{sarazin1986}. 
For this reason, X-rays offer an extremely powerful way to search for galaxy clusters \citep{Vikhlinin1998, bohringer2000, Pierre2016, Adami2018, Koulouridis2021}, and the current all-sky X-ray survey program, extended ROentgen Survey with an Imaging Telescope Array (eROSITA, \citealp{Merloni2012}), are expected to detect a very large sample ($\sim 10^{5}$) of galaxy clusters \citep{Pillepich2012, Clerc2018}. 
More importantly, X-ray observations of galaxy clusters have been proved to be particularly successful in providing information on the distributions of density, temperature, and different metal species within the hot ICM \citep{degrandi2001, Vikhlinin2005, Vikhlinin2006, pratt2007}, which make galaxy clusters powerful probes to study the growth of cosmic structures of the Universe \citep{Allen2011, Kravtsov2012}.

With coeval galaxies are deeply embedded in the gravitational potential wells of the systems, galaxy clusters are also excellent laboratories to study galaxy evolution and its connection with the cluster medium \citep{dressler1980, dressler1997, poggianti1999, rafferty2008}. 
The energy loss of ICM via the radiation of X-ray is expected to form cooling flow in galaxy clusters, which may trigger the star formation processes and thus influence the growth of the galaxies \citep{Fabian1994}.
On the other hand, galaxy evolution feedback, such as stellar activities and the active galactic nuclei (AGN), is thought to have a vital influence on changing the properties of ICM in galaxy clusters \citep{fabian2006,mcnamara2007,bohringer2010,fabian2012,planelles2014}. 

Among all the processes that take place between the galaxy and its surrounding ICM, the AGN-ICM interaction is one of the most spectacular large-scale phenomena that have been observed in galaxy clusters. For example, many clusters of galaxies are found to have prominent cavity structures in their projected X-ray surface brightness images \citep{hlavacek2012, shin2016}, and the X-ray cavities are observed to anti-correlate with the radio emission from the central AGN \citep{boehringer1993, mcnamara2000, fabian2002, birzan2020}. The morphology of the X-ray cavities, together with the flux density contour of the radio lobes, has led to their interpretation as bubbles of relativistic gas blew by the AGN into the thermal ICM \citep{birzan2004, randall2011, birzan2012, fabian2012, mcnamara2012}. 
If the radio galaxy is ploughing through the ICM with a high velocity, ram pressure exerted by the ICM is sufficient to bend jets and sweep the radio-emitting plasma behind the rapidly moving AGN, creating a head-tail (HT) structures in the radio morphology of the radio galaxy \citep{Wellington1973, begelman1979, jones1979}. 
This effect is sensitive to the density of the ICM and HT radio galaxies therefore can be used as tracers of clusters of galaxies \citep{Giacintucci2009, mao2009, Wing2011, garon2019}.

In this work, we report on the discovery of two X-ray luminous galaxy clusters in the direction of the globular cluster NGC 6752. We detect significant X-ray cavities and apparent HT radio galaxies in one of the systems, CXOU J191100-595621, making it one of the excellent targets to study AGN-ICM interactions.  
The paper is organized as follows. In Section 2, we describe our X-ray data analysis, which provides us with X-ray redshifts, temperatures, and metallicities. We also measure the surface brightness profiles and derive the ICM masses and the total cluster masses for the systems. In Section3, we search for optical and radio counterparts of X-ray point sources found in the galaxy clusters. The discussions and conclusions are presented in Section 4. 
Throughout the analysis we adopt the cosmological parameters of $H_{0}=70\, {\rm km\, s^{-1}}\, {\rm Mpc}^{-1}$, $\Omega_{m}=0.3$ and $\Omega_{\Lambda}=0.7$. All quoted uncertainties are given at the $68\%$ confidence level.

\section{Clusters Detection and X-Ray Data Analysis}
\subsection{{\it Chandra} Data and X-Ray Source Detection}

\begin{figure}
	\includegraphics[width=\columnwidth]{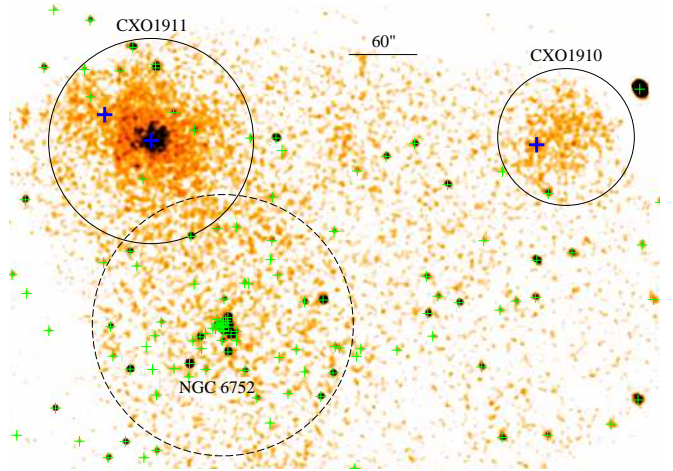}
    \caption{Merged {\it Chandra} 0.5-7.0 keV image of the cluster field, smoothed with a $3\arcsec$ Gaussian kernel. The solid circles represent a radius of $100\arcsec$ for CXO1911 (corresponding to 378 kpc at $z=0.239$) and a radius of $70\arcsec$ for CXO1910 (361 kpc at $z=0.375$). The half-light radius ($R_{h}=1.91\arcmin$) of NGC 6752 is shown as a dashed circle. X-ray point sources are indicated by green pluses, while AGN identified in the galaxy clusters (Section 3) are marked with blue pluses.}
    \label{fig:rawimage}
\end{figure}

The central region of NGC 6752 has been observed by the {\it Chandra} Advanced CCD Imaging Spectrometer (ACIS) 7 times (ObsID: 948, 6612, 19013, 19014, 20121, 20122, 20123), amount to a total effective exposure time of 344 ks. 
Following the standard procedures\footnote{http://cxc.harvard.edu/ciao}, we used the {\it Chandra} Interactive Analysis Observations (CIAO, version 4.11) and the {\it Chandra} Calibration Database (version 4.8.4) to reprocess the data. 
In order to detect the faint X-ray sources in NGC 6752, we aligned and combined all observations into merged images, using the procedures presented in \citet{cheng2019}. 
The two new galaxy clusters, CXOU J191100-595621 (hereafter CXO1911) and CXOU J191012-595619 (hereafter CXO1910), were serendipitously detected in the merged X-ray images of NGC 6752 (Fig-\ref{fig:rawimage}).
Located at a distance of $\sim 2.9\arcmin$ in the north-west direction and a distance of $\sim 5.7\arcmin$ in the north-east direction of NGC 6752, these new systems are immediately classified as extended X-ray sources by visual inspection. Their round X-ray morphologies, high galactic latitudes ($b>25 \degr$), suggest an extragalactic origin rather than come from the Milky Way.
More detailed analysis (Fig-\ref{fig:netcounts}) show that the extended X-ray emission of CXO1911 and CXO1910 are evident in 0.5-7 keV energy bands, with a maximum signal-to-noise ratio of $S/N=43.0$ and $S/N=20.6$, and a maximum net counts of $\sim 7800$ (corresponding to a raiuds of $R\lesssim 100\arcsec$) and $\sim2200$ ($R\lesssim 70\arcsec$), respectively. 

A further study of the X-ray properties of these two clusters is subjected to the contamination of point sources. Therefore, we also search for and create a catalog of X-ray point sources within the merged images, following the steps outlined in \citet{cheng2019}. The details of the catalog will be presented in future work. In this paper, we only focus on exploring the properties of these new diffuse X-ray sources.

\subsection{Spetral Analysis: Redshift, Temperature and Metallicity}
\label{sec:spectra} 

\begin{figure*}
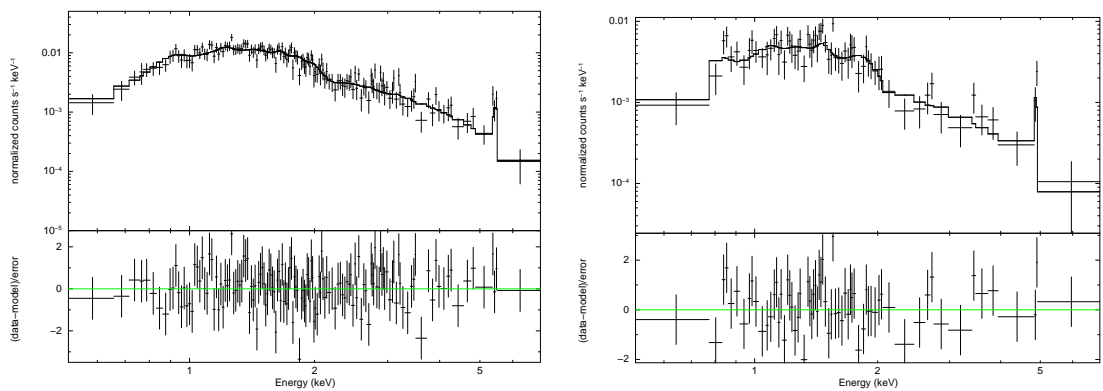

	\includegraphics[width=0.58\columnwidth, angle=-90]{F2a.ps}
	\includegraphics[width=0.58\columnwidth, angle=-90]{F2b.ps}
    \caption{Folded {\it Chandra} spectrum of CXO1911 (left) and CXO1910 (right) with their best-fit mekal models (solid lines). The spectra are binned with at least 20 counts per bin only for visualization, and the lower panels show the residuals. The presence of the iron $K_{\alpha}$ line complex, is evident at $\sim 5.3$ keV (left) and $\sim4.9$ keV (right) in the observed reference frame. We modeled the line complex with a Gaussian model, which gives an equivalent width of $EW\simeq 1.5$ for CXO1911 and $EW\simeq 1.0$ for CXO1910, respectively. }
    \label{fig:spectra}
\end{figure*}

We extract the cluster emission within a circle of radius $R=100\arcsec$ for CXO1911 and $R=70\arcsec$ for CXO1910. As shown in Figure-\ref{fig:netcounts}, these radii are chosen to enclose nearly all of the cluster net source counts within the field of view (FoV). 
Since the defined cluster regions do not always fall within the same CCDs (i.e., S2 and S3 chip of the {\it Chandra} ACIS-S array), we only selected S3 as the chip of spectral extraction, and defined for each observation an exposure-weighted covering factor, which was used to correct the normalization parameter of the spectral fitting.
For the background regions, which are defined as an annulus located on the cluster center, with an inner (outer) radius of $100\arcsec$ ($270\arcsec$) for CXO1911 and $70\arcsec$ ($150\arcsec$) for CXO1910, respectively. We cut out the defined background regions to cover the overlapping field of view of the multiple observations, and ensure that all the background regions are selected in the same chip where the source region lies.
To eliminate the contamination of X-ray point sources, we also excluded the apertures of point sources from the photometry extraction regions. The central region of NGC 6752, is also masked out from the background region of CXO1911. 
Spectral extractions are first performed independently for each observation, we then combined extraction results into a combined spectrum for each cluster.
We analyzed the spectra with XSPEC v12.10.1 \citep{arnaud1996}. A single-temperature {\it mekal} model \citep{kaastra1992,liedahl1995} was used to fit the spectra, in which the ratio of the elements was fixed to the solar value as in \citet{asplund2009}. 
The fits are performed over the energy range 0.5-7.0 keV. We used the Cash statistics \citep{cash1979} applied to the unbinned source plus background counts, and therefore exploit the full spectral resolution of the ACIS instruments. 
Cash statistics also ensure better performance with respect to the canonical $\chi^{2}$ analysis of binned data, particularly for low S/N spectra \citep{nousek1989}.

The best-fit {\it mekal} model gives a neutral hydrogen column density of $N_{\rm H}=(0.29\pm0.08)\times 10^{22} \, {\rm cm^{-2}}$ and a global temperature of $kT=3.32^{+0.57}_{-0.46}$ keV for CXO1911. The measured Fe abundance, in units of \citet{asplund2009}, is $Z_{\rm Fe}=0.64^{+0.34}_{-0.29}Z_{\rm Fe\odot}$, and the unabsorbed rest-frame full-band flux is $S_{0.5-7.0 \, {\rm keV}}=5.36^{+0.72}_{-0.62}\times 10^{-13} {\rm \, erg\, s^{-1}\, cm^{-2}}$. 
For CXO1910, the corresponding best-fit parameters are $N_{\rm H}=0.12^{+0.16}_{-0.12}\times 10^{22} \, {\rm cm^{-2}}$, $kT=3.71^{+1.18}_{-0.86}$ keV, $Z_{\rm Fe}=1.29^{+0.97}_{-0.65}Z_{\rm Fe\odot}$ and $S_{0.5-7.0 \, {\rm keV}}=2.04^{+0.55}_{-0.45}\times 10^{-13} {\rm \, erg\, s^{-1}\, cm^{-2}}$, respectively. 
We find the best-fit column densities are several times higher than the Galactic neutral hydrogen column density measured at the cluster positions (e.g., $N_{\rm H}\sim 6.0\times 10^{20} \, {\rm cm^{-2}}$, \citealp{Willingale2013}). Besides, we also detected significant iron lines from the spectra of the clusters (Figure-\ref{fig:spectra}), and the photon energies are clearly smaller than the rest-frame energy ($\sim 6.7$ keV) of the helium-like Fe $K_{\alpha}$ line complex. These evidence strongly support the galaxy cluster origin of the detected diffuse X-ray emission.

\begin{figure}
	\includegraphics[width=\columnwidth]{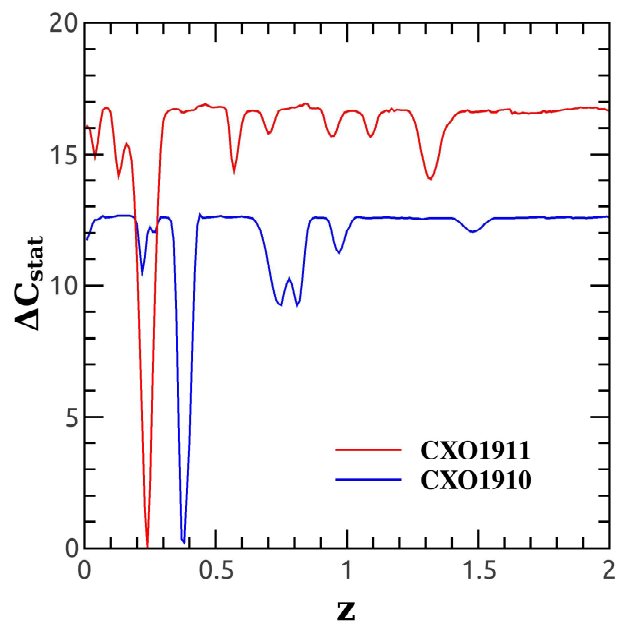}
    \caption{The blind search of $K_{\alpha}$ Fe line in the X-ray spectrum of the clusters. $\Delta C_{\rm stat}$ is shown as a function of the redshift. The minimum is $z=0.239\pm0.013$ and $z=0.375\pm0.016$, with $\Delta C_{\rm stat}\sim 16.7$ and $\Delta C_{\rm stat}\sim 12.6$ for CXO1911 (red) and CXO1910 (blue), respectively. }
    \label{fig:redshift}
\end{figure}

To measure the redshifts of the galaxy clusters, we determined the centroid of the helium-like Fe $K_{\alpha}$ line complex directly from the X-ray analysis. For {\it Chandra} observations, it was estimated that about 1000 net counts are needed to measure $z$ at a $3\sigma$ level, and even more, are needed in the case of hot clusters \citep{yu2011}. Following the method presented by \citet{yu2011}, we ran a combined fit on the ACIS spectra, leaving the neutral hydrogen column density, temperature, metal abundance, redshift, and normalization as free parameters. After finding the absolute minimum, we explore the redshift space by setting other spectral parameters to the best-fit values. The difference of the C-stat value with respect to the minimum is plotted as a function of redshift in Figure-\ref{fig:redshift}. It is obvious that there is a best-fit value of $z=0.239\pm0.013$ for CXO1911, and $z=0.375\pm0.016$ for CXO1910, which corresponds to $\Delta C_{\rm stat}\sim 16.7$ and $\Delta C_{\rm stat}\sim 12.6$ separately. 
As shown in \citet{yu2011}, a larger $\Delta C_{\rm stat}$ indicates a more significant emission line, and a $3\sigma$ confidence level is obtained when $\Delta C_{\rm stat}=9$. Therefore, the value of $\Delta C_{\rm stat}$ suggests that the redshift measurements are highly credible, especially for CXO1911.

\begin{figure*}
	\includegraphics[width=0.9\columnwidth]{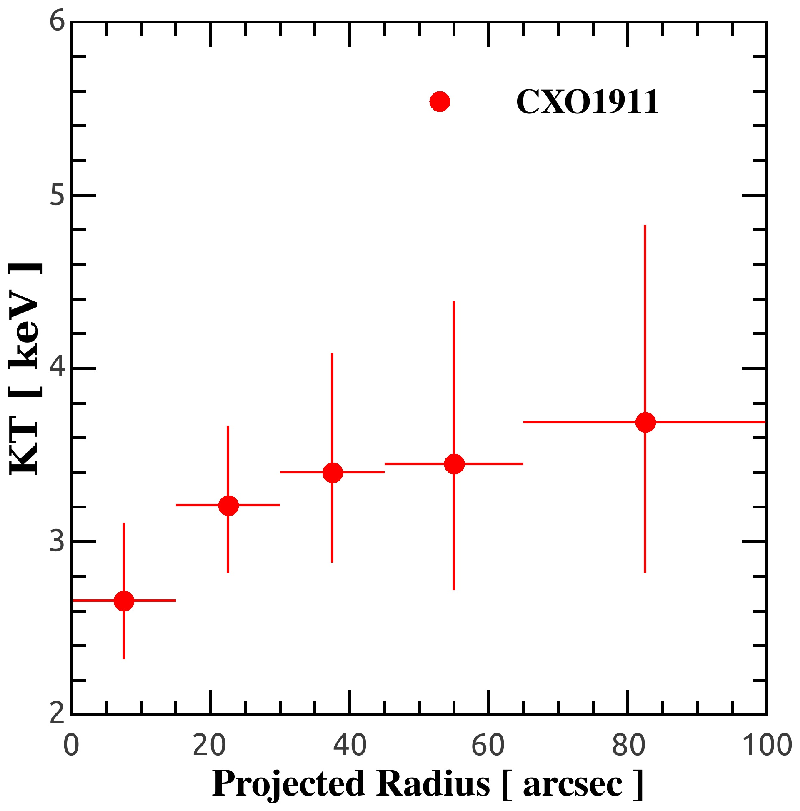}
	\includegraphics[width=0.9\columnwidth]{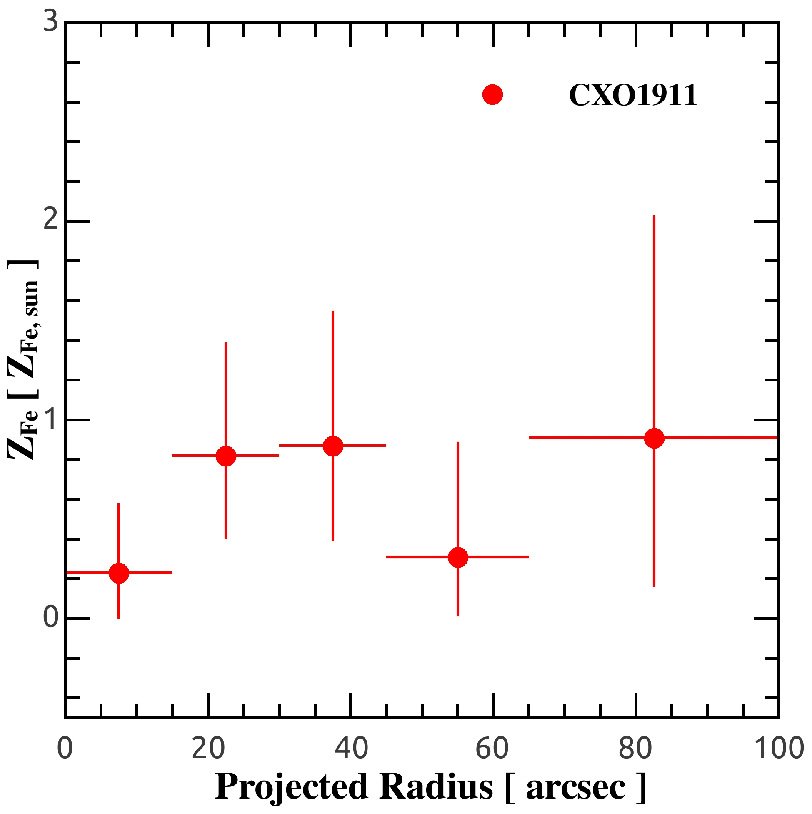}
    \caption{Radial profiles of cluster temperature (left) and metal abundance (right) in CXO1911. The temperature profile decreases with decreasing cluster radius, indicating a cool core.}
    \label{fig:temp_abund_prof}
\end{figure*}

\begin{table*}
	\centering
	\caption{Spectral Analysis Results of the Galaxy Clusters. }
	\label{tab:spectral_fits}
	\begin{tabular}{lcccccr} 
		\hline
		$R$ & $N_{\rm H}$ & kT & $Z_{\rm Fe}$ & $z$ & Norm & $C_{\rm stat}/d.o.f$\\
		($\arcsec$) & ($10^{22} \, \rm cm^{-2}$) & (keV) & ($Z_{\rm Fe\odot}$) &  & ($\times 10^{-4}$) &  \\
		\hline
		\multicolumn{7}{c}{Cluster CXO1910}\\
		\hline
		0-70  & $0.12^{+0.16}_{-0.12}$ & $3.71^{+1.18}_{-0.86}$ & $1.29^{+0.97}_{-0.65}$ & $0.375^{+0.015}_{-0.011}$ & $1.23^{+0.32}_{-0.23}$ & 469.5/439 \\
		\hline
		\multicolumn{7}{c}{Cluster CXO1911}\\
		\hline		
		0-100 & $0.29^{+0.08}_{-0.08}$ & $3.32^{+0.57}_{-0.46}$ & $0.64^{+0.34}_{-0.29}$ & $0.239^{+0.012}_{-0.013}$ & $3.71^{+0.50}_{-0.43}$ & 440.2/439 \\
		0-15  & $0.29$ (frozen) & $2.66^{+0.45}_{-0.34}$ & $0.23^{+0.35}_{-0.23}$ & $0.239$ (frozen) & $0.68^{+0.11}_{-0.10}$ & 496.2/441 \\
	    15-30 & $0.29$ (frozen) & $3.21^{+0.46}_{-0.39}$ & $0.82^{+0.57}_{-0.42}$ & $0.239$ (frozen) & $0.74^{+0.10}_{-0.10}$ & 504.7/441 \\
		30-45 & $0.29$ (frozen) & $3.40^{+0.69}_{-0.52}$ & $0.87^{+0.68}_{-0.48}$ & $0.239$ (frozen) & $0.67^{+0.11}_{-0.10}$ & 454.6/441 \\
		45-65 & $0.29$ (frozen) & $3.45^{+0.94}_{-0.73}$ & $0.31^{+0.58}_{-0.30}$ & $0.239$ (frozen) & $0.85^{+0.19}_{-0.14}$ & 473.6/441 \\
		65-100& $0.29$ (frozen) & $3.69^{+1.14}_{-0.87}$ & $0.91^{+1.12}_{-0.75}$ & $0.239$ (frozen) & $0.81^{+0.23}_{-0.18}$ & 444.4/441 \\
		\hline
	\end{tabular}
\end{table*}

With enough X-ray net counts, we also divided the cluster region of CXO1911 into a set of concentric annuli, in order to study the temperature and metallicity profiles of the cluster. The range of each radial bin was set to contain at least 1000 counts in 0.5-7.0 keV bands. 
We adopted the {\it mekal} model to model the spectrum of each annulus. In all cases, the neutral hydrogen column density and the redshift were fixed at the values obtained in global cluster X-ray spectral fits.
The X-ray spectral fitting results are summarized in Table-\ref{tab:spectral_fits}, and the cluster temperature and metallicity profiles are presented in Figure-\ref{fig:temp_abund_prof}. A slightly cool core is observable in CXO1911.

Finally, we investigate whether our spectral analysis is robust against different choices of the background. Using a ``double subtraction" procedure described in \citet{Li2011} and \citet{cheng2018}, we generated the instrumental background spectra for both the cluster and background regions, using the {\it Chandra} ``stowed" background files. We characterized the local cosmic background spectrum (i.e., instrumental background-subtracted) with an absorbed power-law-plus-Gaussian-lines model. The hence derived cosmic background was added as a fixed component to the total spectral model, after scaling with the exposure-corrected sky area. We find the results of the ``double subtraction" procedure are consistent within $1\sigma$ errors with the analysis based on the direct subtraction of local background. Therefore, we rely only on the spectral analysis obtained with the local background in this paper.

\subsection{X-Ray Surface Brightness Distribution}

To measure the surface brightness of the clusters we use only the merged exposure-corrected images. Considering that the calculation of cluster radial brightness profile is subject to the contamination of X-ray point sources, we also masked out the regions of all point sources from the merged images, and then filled in the masked region for each source with background counts from a local annulus. We detected an X-ray point source (CXOU J191100.49-595621.5, $\alpha=19^{h}11^{m}00^{s}.5$ and $\delta=-59\degr 56\arcmin 21.5\arcsec$) at the geometrical diffuse X-ray emission center of CXO1911, which is overlapped with a radio galaxy (see next section) and thus is identified to be the central AGN of CXO1911. For CXO1910, no significant point source was detected near the geometrical center. We therefore calculate a photon bevy center of $\alpha=19^{h}10^{m}12^{s}.6$ and $\delta=-59\degr 56\arcmin 19\arcsec$ for this cluster.

\begin{figure}
	\includegraphics[width=\columnwidth]{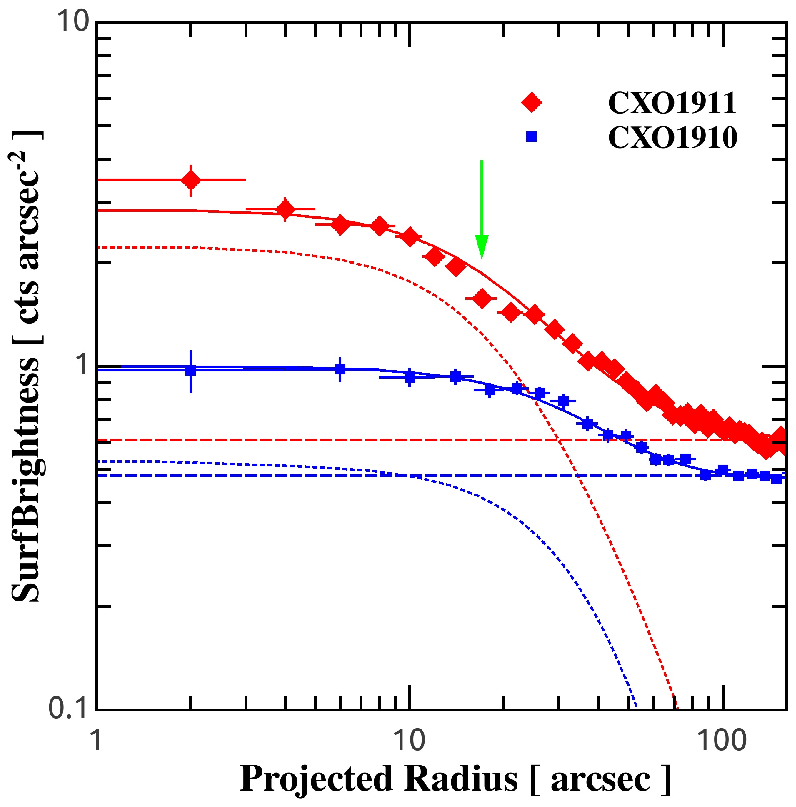}
    \caption{Surface brightness profiles of clusters in 0.5-7.0 keV bands, red for CXO1911 and blue for CXO1910. The best-fitting $\beta$ models, $S(r)=S_{0}(1+(R/R_{c})^{2})^{-3\beta+0.5}+{\rm bkg}$, are shown as solid lines, while the cluster components ($S_{0}(1+(R/R_{c})^{2})^{-3\beta+0.5}$) and the background levels (${\rm bkg}$), are presented as dotted curves and horizontal dashed lines, respectively. Green arrow marks the location ($R\sim 17\arcsec$) of the X-ray cavities in CXO1911 (Figure-\ref{fig:cavities}).  }
    \label{fig:surf_brightness}
\end{figure}

\begin{figure}
	\includegraphics[width=\columnwidth]{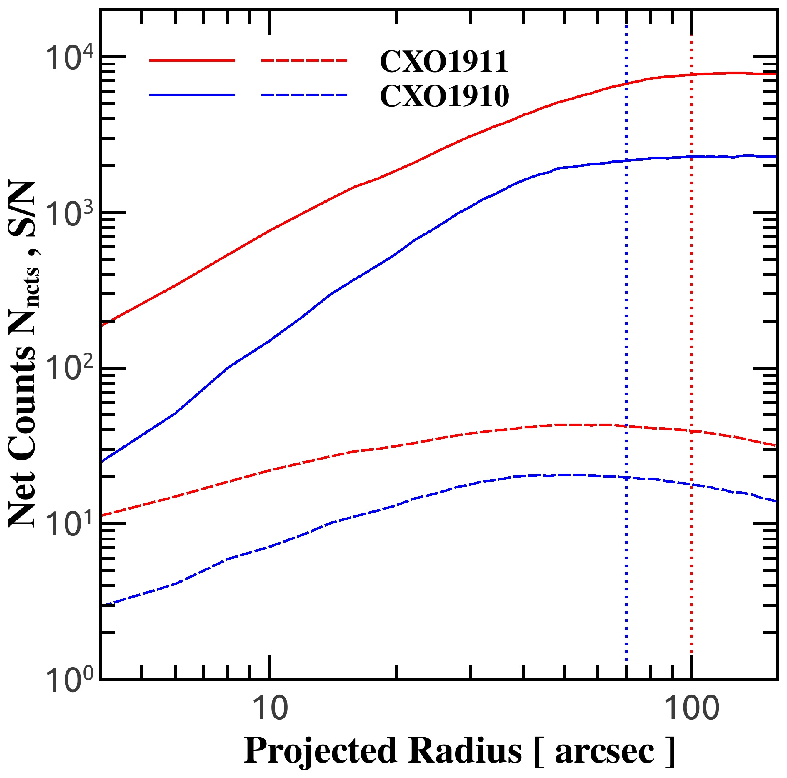}
    \caption{Cluster net source counts (solid lines) and signal-to-noise ratio $S/N$ (dashed lines) within a given projected radius. We masked out the detected point sources from the merged images, and calculated the net source counts $N_{ncts}$ and signal-to-noise ratio $S/N$ with the background levels obtained in Figure-\ref{fig:surf_brightness}. The data are calculated in 0.5-7.0 keV bands, with red and blue curves marks CXO1911 and CXO1910, respectively. Vertical dotted lines represent the $N_{ncts}$ saturated radius of the clusters. }
    \label{fig:netcounts}
\end{figure}

The azimuthal averaged surface brightness profiles are shown as red dots and blue squares in Figure-\ref{fig:surf_brightness}, which are well described by the $\beta$ model \citep{cavaliere1976}, i.e.
\begin{equation}
 S(r)=S_{0}(1+(R/R_{c})^{2})^{-3\beta+0.5}+{\rm bkg}.
	\label{eq:beta_model}
\end{equation}
Using a maximum likelihood algorithm, we fit the surface brightness profile of each cluster with the $\beta$ model. The best-fit model gives a slope of $\beta=0.55\pm0.06$ and a core radius of $R_{c}=(20.9\pm2.4)\arcsec$ for CXO1911, and $\beta=0.89^{+0.51}_{-0.32}$ and $R_{c}=(51.9^{+19.0}_{-15.8})\arcsec$ for CXO1910, respectively. Here the fitting confidence is at the $1\sigma$ level.
From Figure-\ref{fig:surf_brightness}, it can be seen that the cluster components (doted lines) decrease quickly outside of the core radius, which become an order of magnitude lower than the background levels (horizontal dashed lines) at $R\sim 100\arcsec$ (for CXO1911) and $R\sim 70\arcsec$ (for CXO1910). We calculated the net counts ($N_{ncts}$) of the clusters with the best-fitted background levels ($bkg$). The cumulative distributions suggest that $N_{ncts}$ increases monotonically with increasing $R$, which become saturated around the cluster radius (i.e., $R\sim 100\arcsec$ for CXO1911 and $R\sim 70\arcsec$ for CXO1910). While beyond the cluster radius, the values of $N_{ncts}$ are nearly constants, with $N_{ncts}\approx 7800$ for CXO1911 and $N_{ncts}\approx 2200$ for CXO1910, respectively (Figure-\ref{fig:netcounts}). 

\begin{figure*}
	\includegraphics[width=2.0\columnwidth]{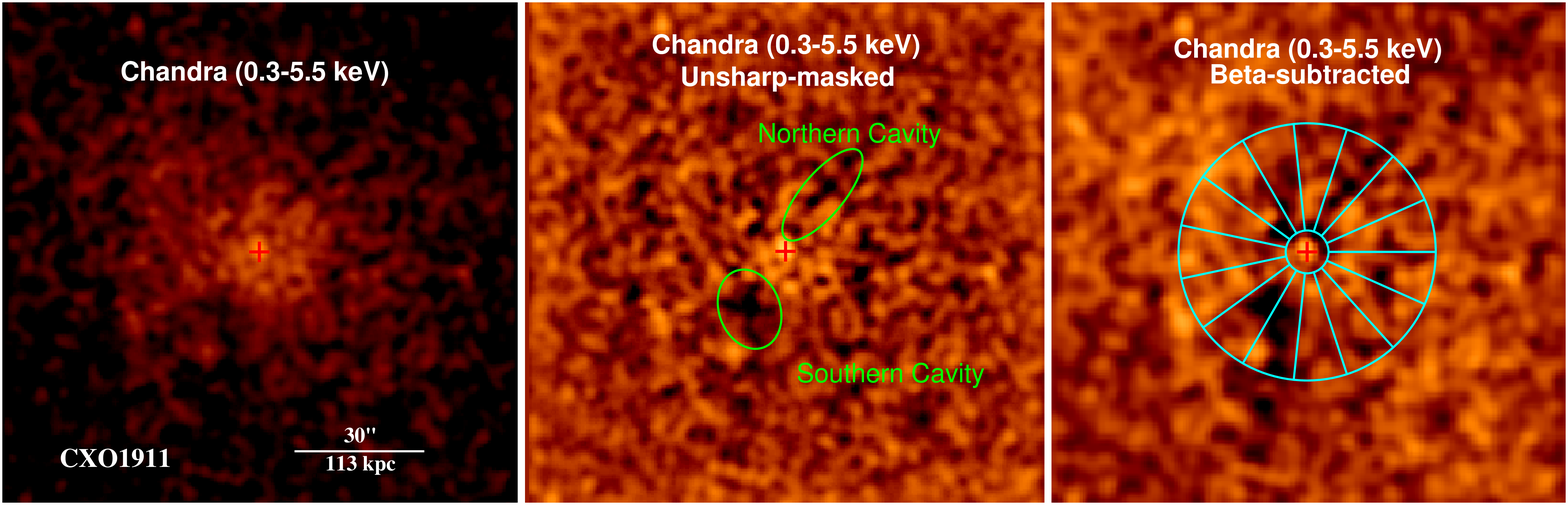}
    \caption{Merged and lightly smoothed {\it Chandra} image (left), unsharp-masked image (middle) and $\beta$-subtracted image (right) for CXO1911. We highlight the X-ray cavities with green ellipses. The position of the central AGN, taken to be the position of the BCG, is shown with the red cross. In cyan, we illustrate the annuli employed to compute the azimuthal surface brightness profiles (Figure-\ref{fig:azimuthal_brightness}). }
    \label{fig:cavities}
\end{figure*}

\begin{figure}
	\includegraphics[width=\columnwidth]{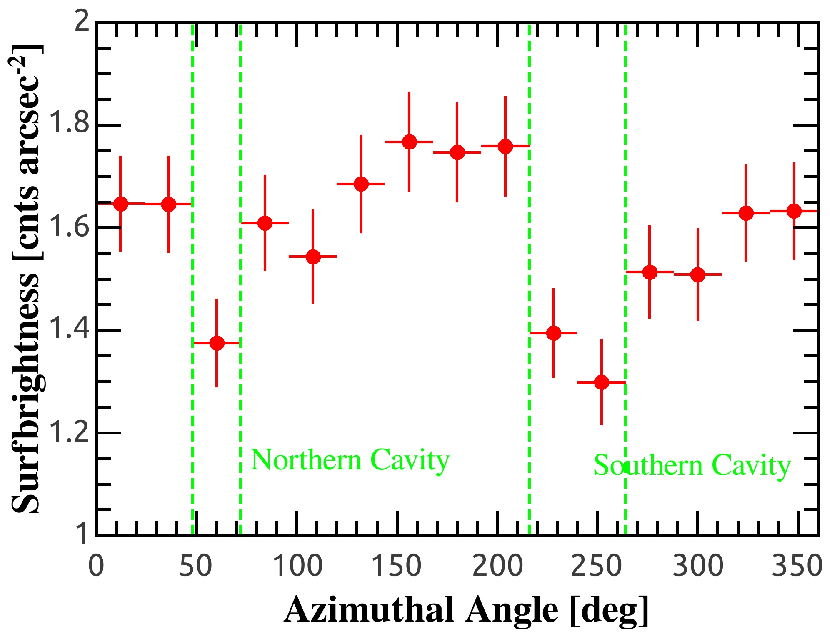}
    \caption{{\it Chandra} 0.3-5.5 keV azimuthal surface brightness profile of CXO1911. The profile is calculated with the annulus illustrated in the right panel of Figure-\ref{fig:cavities}, which was divided into 15 sectors. Sectors occupied by the X-ray cavities are marked with vertical green lines. The counter-clockwise zero angles corresponds to the right direction in Figure-\ref{fig:cavities}.  }
    \label{fig:azimuthal_brightness}
\end{figure}

Through visual inspection, we also find significant X-ray cavities in the surface brightness images of CXO1911. To better demonstrate the features of the cavities, we use only the 0.3-5.5 keV band images, which have higher $S/N$ than the full band images. As a first indicator, we created an unsharp-masked image to enhance the deviations of the X-ray cavities. As suggested by \citet{blanton2009}, this method consists of subtracting a strongly smoothed image from a lightly smoothed image and has been used extensively in the literature for X-ray cavity studies (e.g., \citealp{sanders2009,machacek2011,hlavacek2015}). For the lightly smoothed image, we use a Gaussian smoothing scale of 10 kpc ($\sim 2.5\arcsec$) to match the length scale of the cavities, and for the strongly smoothed image, we use a Gaussian smoothing scale of 60 kpc ($\sim 16\arcsec$) to match the underlying large-scale cluster emission. Besides, we also build a $\beta$ model with the best-fitting function obtained in Figure-\ref{fig:surf_brightness}. We subtracted the $\beta$ model from the lightly smoothed image and refer the resulting image as the "$\beta$-subtracted" image. As shown in \citet{hlavacek2015}, these images are helpful in identifying possible X-ray cavities from galaxy clusters.
The lightly smoothed image, unsharp-masked image, and $\beta$-subtracted image are presented in Figure-\ref{fig:cavities}. A pair of symmetric surface brightness cavities (green ellipses) is evident in the figures, which have tubular structures and are connected with the AGN at the cluster center. To quantify the significance of the cavities, we measure azimuthal surface brightness profiles from the X-ray data (left panel of Figure-\ref{fig:cavities}), using an annulus that encompassed the cavities ($5\arcsec \leq R \leq 30\arcsec$), centered on the X-ray peak (right panel of Figure-\ref{fig:cavities}). The annulus was divided into 15 azimuthal sectors, containing roughly 300 X-ray counts per sector. The sectors of the cavities, show a counts depressions of $\sim 15\%-20\%$, which are $\sim 2.5\sigma-3\sigma$ smaller than other sectors (Figure-\ref{fig:azimuthal_brightness}). According to the criteria of X-ray cavity definition of nearby clusters, these cavities belong to the ``clear'' X-ray cavities (with X-ray surface brightness depressions lie in $\sim 2\sigma-3\sigma$ (or more) and contain $\sim 10\%-40\%$ fewer counts) and are consistent with cavities seen in local clusters \citep{hlavacek2015}.

\subsection{Mass Estimate}

The baryonic mass ($M_{\rm ICM}$) can be directly measured once the three-dimensional (3D) electron density profile is known. For the projected surface density $\beta$-model, which can be deprojected into $n_{e}(r)=n_{e0}[(1+(r/r_{c})^{2})^{-3\beta/2}]$, with $n_{e0}$ the central 3D electron density. To measure $n_{e0}$, we 
use the relation between the normalization of the X-ray spectrum and the electron ($n_{e}$) and hydrogen ($n_{H}$) density in the intra cluster medium (ICM) for the {\it mekal} model:
\begin{equation}
    {\rm Norm}=\frac{10^{-14}}{4\pi {[D_{a}(1+z)]}^{2}}\times \int n_{e}n_{H}dV,
	\label{eq:norm}
\end{equation}
where $D_{a}$ is the angular diameter distance to the source (${\rm cm}$), $z$ is the redshift, $n_{H}=0.85n_{e}$ (${\rm cm^{-3}}$), and the volume integral is performed over the projected region used for the spectral fit. 
For the global spectral fits (i.e., $0\arcsec \leq R\leq 100\arcsec$ for CXO1911 and $0\arcsec \leq R\leq 70\arcsec$ for CXO1910) performed in Table-\ref{tab:spectral_fits}, this gives a central density of $n_{e0}=(5.37\pm0.61)\times 10^{-3}\, {\rm cm^{-3}}$ and $n_{e0}=(1.79\pm0.26)\times 10^{-3}\, {\rm cm^{-3}}$ for CXO1911 and CXO1910, respectively. 

Assuming that the clusters are in hydrostatic equilibrium and their temperature and density distributions, as well as the total gravitational potential, are spherically symmetric, we can also estimate the total gravitating mass of the clusters with equation \citep{sarazin1988}:
\begin{equation}
    M(<r)=-4.0\times 10^{13}M_{\odot}T(r)r(\frac{d{\rm log}\,n_{e}}{d{\rm log}\,r}+\frac{d{\rm log}\,T(r)}{d{\rm log}\,r}).
	\label{eq:mass}
\end{equation}
Here $M(<r)$ is the estimated total gravitating mass within a cluster-centric distance $r$ (in units of Mpc), and $T(r)$ is the 3D gas temperature (in units of keV) at the radius $r$. Since the temperature profiles are not well measured in these two clusters, we assume isothermality inside the spectral extraction radius and beyond that adopt a mildly decreasing temperature profile $T \propto r^{-0.24}$, as found in local clusters \citep{leccardi2008}.
Therefore the term of parenthesis in Equation-\ref{eq:mass} can be simply written as $-3\beta x^{2}/(1+x^{2})$, where $x=r/r_{c}$. Furthermore, considering that the slope of the density profile is robustly measured only within the cluster extraction region (i.e., $R=378$ kpc for CXO1911 and $R=361$ kpc for CXO1910), we choose to keep the slope constant (1.58 for CXO1911 and 1.72 for CXO1910) beyond the extraction radius. 

\begin{table}
	\centering
	\caption{Summary of the Mass Estimates for the Clusters. }
	\label{tab:mass_estimates}
	\begin{tabular}{ccc} 
		\hline
	    Radius & $M_{\rm ICM}$ & $M_{\rm tot}$ \\
		(kpc)  & ($M_{\odot}$) & ($M_{\odot}$) \\
		\hline
		\multicolumn{3}{c}{Cluster CXO1911}\\
		\hline
		$r_{c}=79\pm9.6$     & $(2.2\pm0.3)\times 10^{11}$ & $(8.7\pm2.5)\times 10^{12}$ \\
		$r_{2500}=340\pm34$  & $(4.0\pm0.8)\times 10^{12}$ & $(7.1\pm2.1)\times 10^{13}$ \\
		$r'=378$              & $(4.7\pm1.0)\times 10^{12}$ & $(7.9\pm2.4)\times 10^{13}$ \\
		$r_{500}=705\pm72$   & $(1.2\pm0.3)\times 10^{13}$ & $(1.3\pm0.4)\times 10^{14}$ \\	
        $r_{200}=1060\pm125$ & $(2.2\pm0.6)\times 10^{13}$ & $(1.7\pm0.6)\times 10^{14}$ \\		
		\hline
		\multicolumn{3}{c}{Cluster CXO1910}\\
		\hline
		$r_{c}=268\pm98$      & $(2.3\pm0.7)\times 10^{12}$ & $(5.3\pm2.7)\times 10^{13}$ \\		
		$r_{2500}=340\pm69$   & $(3.7\pm1.5)\times 10^{12}$ & $(8.3\pm5.0)\times 10^{13}$ \\
		$r'=361$               & $(4.2\pm1.7)\times 10^{12}$ & $(9.2\pm5.8)\times 10^{13}$ \\
		$r_{500}=780\pm195$   & $(1.3\pm0.7)\times 10^{13}$ & $(2.0\pm1.5)\times 10^{14}$ \\	
        $r_{200}=1230\pm359$  & $(2.1\pm1.4)\times 10^{13}$ & $(3.1\pm2.8)\times 10^{14}$ \\
		\hline
	\end{tabular}
\end{table}

The baryonic mass ($M_{\rm ICM}$) and total mass ($M_{\rm tot}$) is calculated within cluster radius ($r'$) and core radius ($r_{c}$) separately, and the results are listed in Table-\ref{tab:mass_estimates}. For comparison, we also estimated $M_{\rm tot}$ for each cluster at three typical density contrast levels (i.e., $\Delta=2500$, 500 and 200) with respect to the critical density $\rho_{c}(z)$. The corresponding cluster-centric distances ( $r_{\Delta}$) is derived by solving the equation $M_{\Delta}(r_{\Delta})=4\pi\Delta r_{\Delta}^{3}\rho_{c}(z)/3$. The $1\sigma$ confidence intervals are computed by including the error on the temperature and on the gas density profiles. The mass estimation results are reported in Table-\ref{tab:mass_estimates}. We find only the radius $r_{2500}$ is well within the cluster extraction region, which corresponds to 
a total mass of $M_{2500}=(7.1\pm2.1)\times10^{13}\,M_{\odot}$ for CXO1911, and $M_{2500}=(8.3\pm5.0)\times10^{13}\,M_{\odot}$ for CXO1910. The ICM mass fraction is $f_{\rm ICM,2500}=0.057\pm0.031$ for CXO1911 and $f_{\rm ICM,2500}=0.045\pm0.032$ for CXO1910, respectively, which are consistent with the values (i.e., $f_{\rm ICM,2500}\sim0.02-0.1$, \citealp{sun2009}) measured in nearby galaxy clusters.

The total masses extrapolated to $r_{500}$, with the additional assumption about the temperature and density profiles, are $M_{500}=(1.3\pm0.4)\times10^{14}\,M_{\odot}$ for CXO1911 and $M_{500}=(2.0\pm1.5)\times10^{14}\,M_{\odot}$ for CXO1910. We also extrapolate the mass measurement up to the nominal virial radius $r_{200}$, finding $M_{200}=(1.7\pm0.6)\times10^{14}\,M_{\odot}$ for CXO1911 and $M_{200}=(3.1\pm2.8)\times10^{14}\,M_{\odot}$ for CXO1910. To evaluate the possible systematics associated with the extrapolation of the total masses, we present in Table-\ref{tab:empirical_calibrations} the mass estimates adopting two different empirical calibration relations. The first estimate is based on the $M_{500}-T$ scaling relation, which is calibrated on local clusters and presented in \citet{Vikhlinin2009}. The second is the integrated Compton parameter $Y_{X}\equiv T_{\rm X}\times M_{\rm ICM}$, which is considered a robust mass proxy within $r_{500}$ \citep{arnaud2010}.  
According to Table-\ref{tab:mass_estimates} and Table-\ref{tab:empirical_calibrations}, it is clear that all the mass estimates are consistent well within $1\sigma$ with each other.

\begin{table}
	\centering
	\caption{Total Mass Estimates of the Clusters Obtained through Empirical Relations. }
	\label{tab:empirical_calibrations}
	\begin{tabular}{@{}lll@{}} 
		\hline
		\multicolumn{1}{l}{\multirow{2}{*}{Method}} & \multicolumn{2}{c}{$M_{500}$ ($M_{\odot}$)}\\
		\cline{2-3}
	    \multicolumn{1}{c}{}           &\multicolumn{1}{c}{CXO1911}    & \multicolumn{1}{c}{CXO1910}  \\
		\hline
	    (1): $M_{500}-T$      & $(1.4\pm0.4)\times 10^{14}$ & $(1.6\pm0.8)\times 10^{14}$ \\
		(2): $Y_{X}-M_{500}$  & $(1.3\pm0.2)\times 10^{14}$ & $(1.4\pm0.6)\times 10^{14}$ \\		
		\hline
	\end{tabular}
	
{\bf Notes.} Empirical calibration relation from (1): \citet{Vikhlinin2009}; (2): \citet{arnaud2010}. \hfill
\end{table}

\section{The Multiband Counterparts}
\subsection{Identification of Galaxy Members in CXO1911}
 
\begin{figure*}
	\includegraphics[width=2.0\columnwidth]{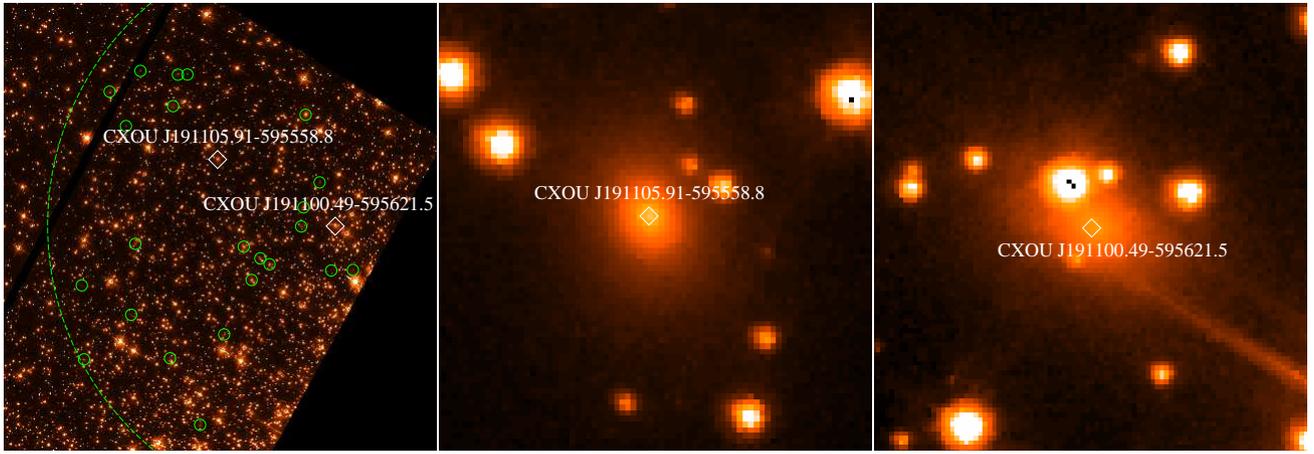}
    \caption{Left panel: {\it HST} ACS/F606W $150\arcsec \times 150\arcsec$ image of CXO1911. Galaxies identified by visual inspection are shown as green circles, and white diamonds mark the galaxies that have an active nucleus (i.e., galaxies with a counterpart X-ray point source in Figure-\ref{fig:rawimage}). The green dashed curve represents the cluster region ($R=100\arcsec$) of CXO1911. Middle panel: the zoomed-in image of CXOU J191105.91-595558.8, which is found to be a narrow-angle-tailed radio galaxy (Figure-\ref{fig:radio_img}). Right panel: the zoomed-in image of the brightest central galaxy (BCG) of CXO1911, which is identified to be associated with the X-ray point source CXOU J191100.49-595621.5. Black dots in the middle and right panels are the pileup pixels of the foreground bright stars. }
    \label{fig:optical_img}
\end{figure*}

Because of the proximity to NGC 6752, the FoVs of CXO1911 and CXO1910 are strongly contaminated by the interloper stars from the foreground globular cluster, which hampers the identification of galaxy members in these two systems. Indeed, we have queried the NASA/IPAC Extragalactic Database and SIMBAD Astronomical Database for possible galaxies, and no entries of galaxies are retrieved within the FoVs of these two systems. Fortunately, NGC6752 is one of the globular clusters that has been frequently observed by the {\it Hubble Space Telescope (HST)}, which enable us to identify the galaxy members with superb angular resolution and sensitivity, especially with the images collected by the Wide Field Channel (WFC) of the Advanced Camera for Surveys (ACS). 
For CXO1911, we find its FoVs is partially covered by one of the ACS/WFC field of {\it HST} in program HAP-10458 (PI:Biretta), and tens of extended sources can be clearly identified as galaxies by visual inspection (Figure-\ref{fig:optical_img}). We cross-matched the {\it Chandra} X-ray point sources with the detected galaxies. Two point sources, including CXOU J191100.49-595621.5 and CXOU J191105.91-595558.8, are found to have a counterpart elliptical galaxy, suggesting that the X-ray sources are AGN of the host galaxies. Located in the X-ray emission center of CXO1911, the counterpart galaxy of CXOU J191100.49-595621.5 is also found to have the smallest visual magnitude $m_{\rm F606W}$ among all detected galaxies. Therefore, we identified this galaxy as the brightest cluster galaxy (BCG) of CXO1911. Since no {\it HST} images are found to cover the FoVs of CXO1910, we found no counterpart galaxies in this galaxy cluster.

\subsection{Radio Counterparts}

\begin{table*}
	\centering
	\caption{Multiband Counterparts of AGN candidates in CXO1911 and CXO1910. }
	\label{tab:counterparts}
	\begin{tabular}{@{}ccccc@{}} 
		\hline
		\multicolumn{1}{l}{\multirow{2}{*}{Parameters}} & \multicolumn{2}{c}{CXO1911} & \multicolumn{1}{c}{} &  \multicolumn{1}{c}{CXO1910}\\
		\cline{2-3}\cline{5-5}
	    \multicolumn{1}{c}{}  &\multicolumn{1}{c}{CXOU J191100.49-595621.5} & \multicolumn{1}{c}{CXOU J191105.91-595558.8} & \multicolumn{1}{c}{} & \multicolumn{1}{c}{CXOU J191015.43-595624.9} \\
		\hline
X-ray Coordinates & ($19^{h}11^{m}00^{s}.49$, $-59\degr 56\arcmin 21\arcsec.5$) & ($19^{h}11^{m}05^{s}.91$, $-59\degr 55\arcmin 58\arcsec.8$) & & ($19^{h}10^{m}15^{s}.43$, $-59\degr 56\arcmin 24\arcsec.9$) \\		
$S_{\rm 0.5-8 \,keV}$ (${\rm erg\,s^{-1}\,cm^{-2}}$) & $(9.73\pm2.75)\times 10^{-16}$ & $(1.09\pm0.32)\times 10^{-15}$ & &  $(1.94\pm0.51)\times 10^{-15}$  \\	
{\it HST} Galaxy Coordinates & ($19^{h}11^{m}00^{s}.66$, $-59\degr 56\arcmin 21\arcsec.88$) & ($19^{h}11^{m}06^{s}.05$, $-59\degr 55\arcmin 59\arcsec.00$) & & --- \\
$m_{\rm F606W}$  & $19.191\pm2.237$ & $19.911\pm1.939$ & & --- \\
170--231 MHz Coordinates & ($19^{h}11^{m}08^{s}$, $-59\degr 55\arcmin 56\arcsec$) & ($19^{h}11^{m}08^{s}$, $-59\degr 55\arcmin 56\arcsec$) & & ($19^{h}10^{m}15^{s}$, $-59\degr 56\arcmin 15\arcsec$) \\
$S_{\rm 170-231\, MHz}$ (mJy) & $327.6\pm26.2$ & $327.6\pm26.2$ & & $146.4\pm11.7$ \\
843 MHz Coordinates & ($19^{h}10^{m}59^{s}.81$, $-59\degr 56\arcmin 23\arcsec.3$) & ($19^{h}11^{m}08^{s}.5$, $-59\degr 55\arcmin 50\arcsec.2$) & & ($19^{h}10^{m}15^{s}.29$, $-59\degr 56\arcmin 22\arcsec.5$) \\
$S_{\rm 843\, MHz}$ (mJy)     & $25.3\pm1.2$ & $66.8\pm3.4$ & & $54.0\pm1.9$\\
5.5 GHz Coordinates & ($19^{h}11^{m}00^{s}.50$, $-59\degr 56\arcmin 21\arcsec.5$) & ($19^{h}11^{m}05^{s}.89$, $-59\degr 55\arcmin 58\arcsec.4$) & & ($19^{h}10^{m}15^{s}.80$, $-59\degr 56\arcmin 21\arcsec.71$) \\
$S_{\rm 5.5\, GHz}$ (mJy)     & $20.0\pm2.0$ & $2.1\pm0.2$ & & $18.4\pm1.8$\\
9 GHz Coordinates & ($19^{h}11^{m}00^{s}.51$, $-59\degr 56\arcmin 21\arcsec.6$) & ($19^{h}11^{m}05^{s}.89$, $-59\degr 55\arcmin 58\arcsec.7$) & & ($19^{h}10^{m}15^{s}.80$, $-59\degr 56\arcmin 21\arcsec.71$) \\
$S_{\rm 9\, GHz}$ (mJy)     & $17.4\pm1.7$ & $0.1\pm0.1$ & & $0.4\pm0.1$\\

		\hline
	\end{tabular}
\end{table*}

\begin{figure*}
	\includegraphics[width=0.95\columnwidth]{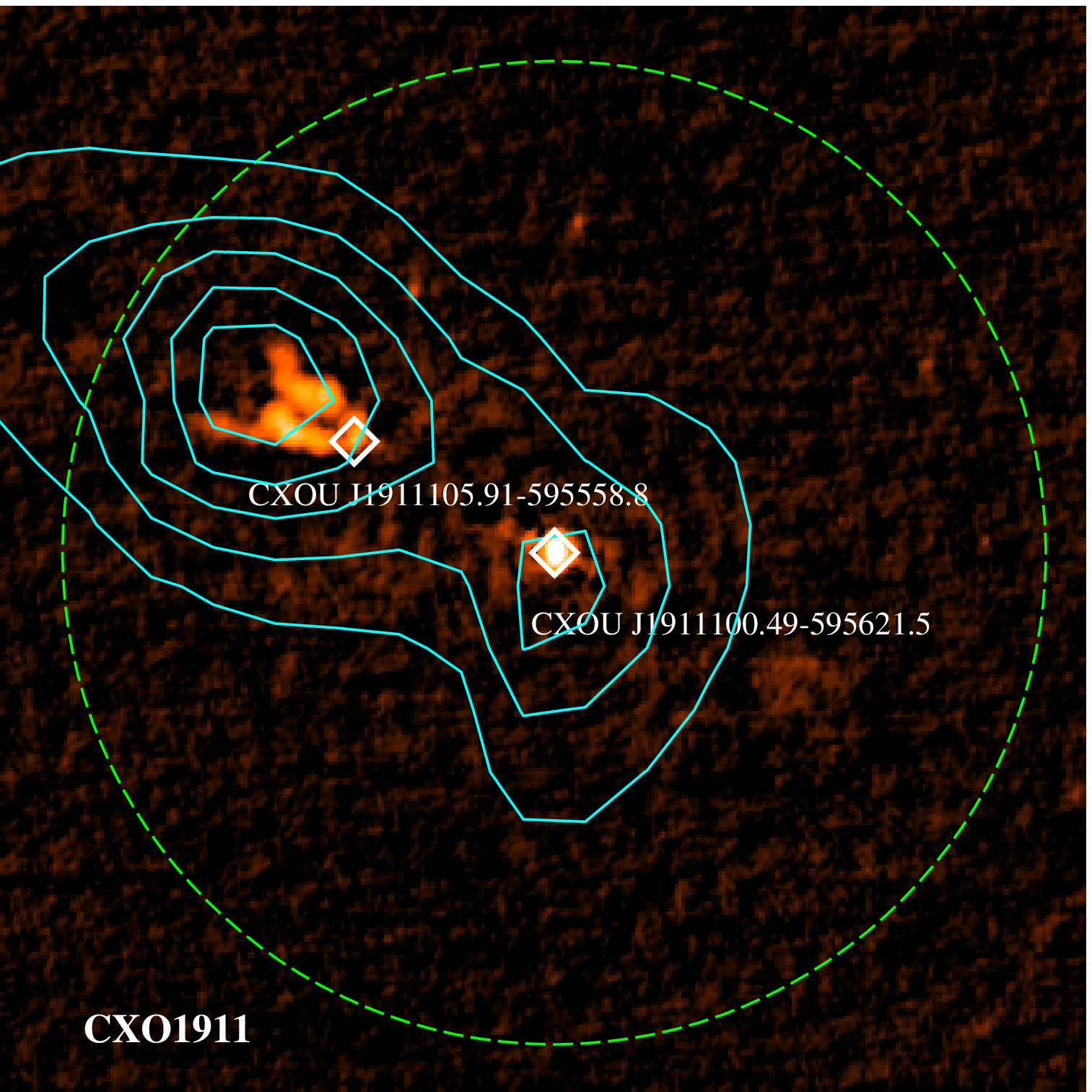}
	\includegraphics[width=0.95\columnwidth]{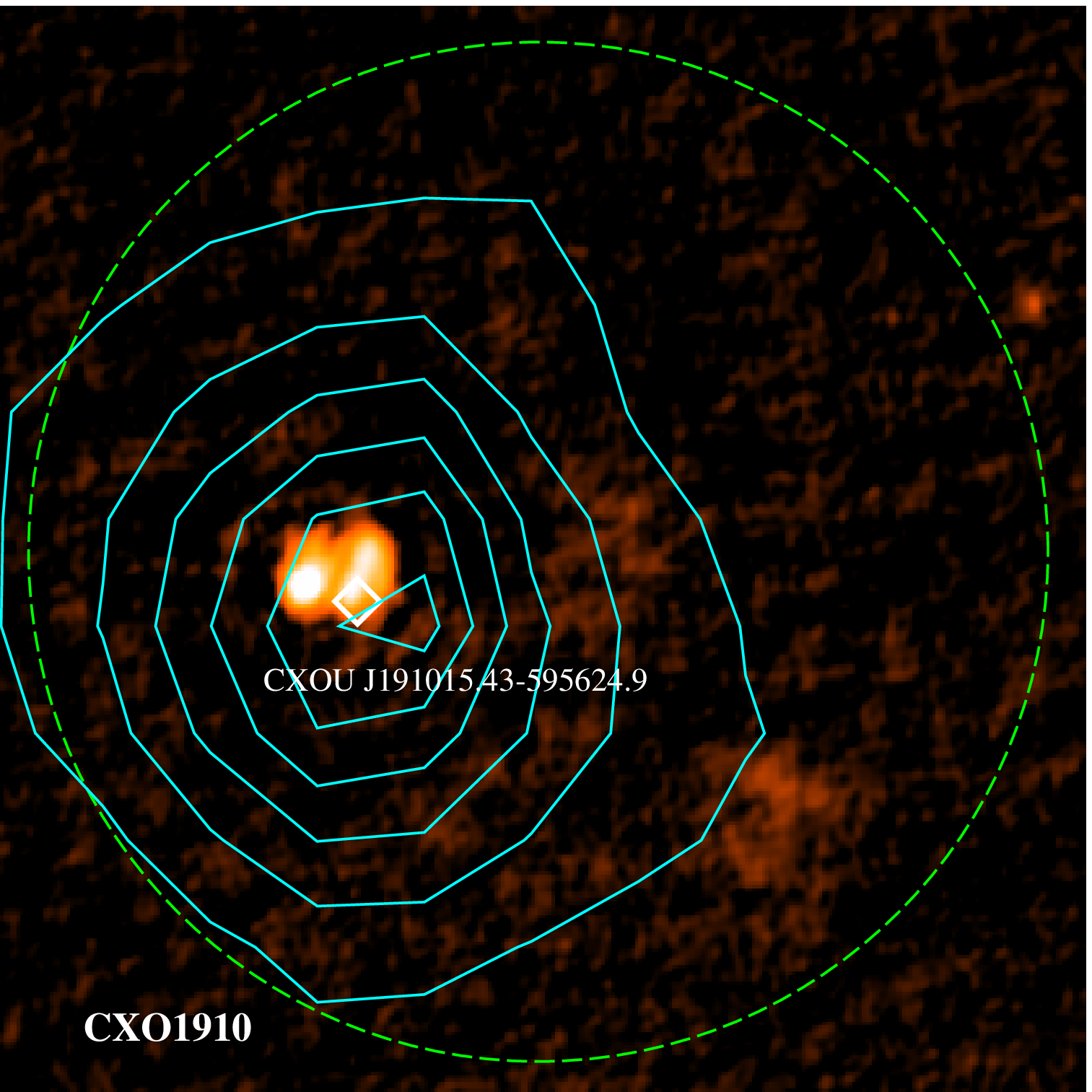}
    \caption{ATCA 5.5 GHz intensity maps of CXO1911 (left) and CXO1910 (right) with the SUMSS 843 MHz contours (cyan curves) overlaid. X-ray point sources associated with optical galaxies or radio emissions are shown as white diamonds. CXOU J191105.91-595558.8 is identified to be a narrow-angle-tail radio galaxy with a pair of symmetric and elongated radio tails. The green dashed circles represent the cluster region with a radius of $R=100\arcsec$ for CXO1911 and $R=60\arcsec$ for CXO1910, respectively.}
    \label{fig:radio_img}
\end{figure*}

At 72-231 MHz, CXO1911 and CXO1910 are found to be associated with two radio sources, J191108-595556 and J191015-595615, detected by the Galactic and Extragalactic All-sky MWA survey (GLEAM, \citealp{hurley2017}). 
At higher frequency, we find the morphology of J191108-595556 splits into two parts according to the 843 MHz images of the Sydney University Molonglo Sky Survey (SUMSS, \citealp{mauch2003}), and the coordinate of the southeastern part is roughly coincident with the X-ray emission center of CXO1911. However, because of the resolution of the GLEAM ($\sim 2\arcmin/{\rm beam}$) and SUMSS ($\sim 45\arcsec/{\rm beam}$) images are comparable to the angular size of the galaxy clusters, the nature of the radio sources is far from convincing. 
Fortunately, the FoVs of CXO1911 and CXO1910 have been surveyed by the Australia Telescope Compact Array (ATCA) in February 2014, with a total observation time of $\sim 20$ hours and a 2 GHz bandwidth at 5.5 GHz and 9.0 GHz (PI: J. Strader). This offers us an opportunity to resolve the radio morphology with much higher resolution ($\sim 2.6\arcsec/{\rm beam}$ and $\sim 1.6\arcsec/{\rm beam}$) and sensitivity.

The ATCA 5.5 GHz intensity maps are presented in Figure-\ref{fig:radio_img}. The southeastern part of J191108-595556 is resolved to be a point source, which is coincident with CXOU J191100.49-595621.5 (with superposition accuracy better than $0.5\arcsec$) and suggesting a common AGN origin of the X-ray and radio emission from the optical BCG. The northwestern part of J191108-595556 is found to be more complex, with a semicircular shape of the radio head that orientates the center of the galaxy cluster, a deep well behind the head, and a pair of symmetric radio lobes which stretch far down the tail. The radio head is associated with CXOU J191105.91-595558.8 and the optical galaxy, with superposition accuracies better than $\sim 1\arcsec$. These features suggest that CXOU J191105.91-595558.8 is a narrow-angle-tail (NAT) radio galaxy, and the elongated radio tails can be explained as trails left as the radio active galaxy plough through the ICM of the galaxy cluster. 
For CXO1910, the morphology of J191015-595615 is found to split into two extended parts in the 5.5 GHz and 9.0 GHz images. A {\it Chandra} point source, CXOU J191015.43-595624.9, is found to be roughly coincident with the eastern part of the radio morphology, indicating that this source is an AGN candidate of this galaxy cluster.  

\begin{figure}
	\includegraphics[width=\columnwidth]{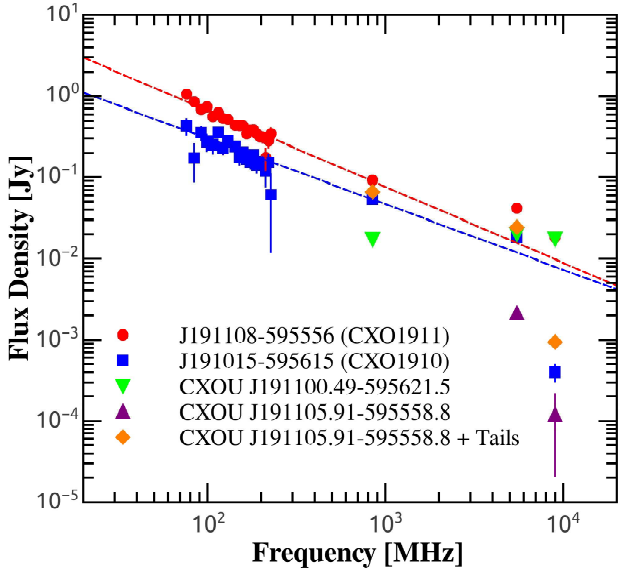}
    \caption{Radio spectra of the candidate radio galaxies between 70 MHz and 9.0 GHz. Red dots and blue squares represent the total intensity of CXO1911 and CXO1910, respectively. At 72-231 MHz, CXO1911 and CXO1910 are found to be associated with the GLEAM sources J191108-595556 and J191015-595615. At 843 MHz, J191108-595556 splits into two sources, CXOU J191100.49-595621.5 (green down triangles) and a NAT radio galaxy (yellow diamonds), with CXOU J191105.91-595558.8 (purple up triangles) is identified to be the radio head of the NAT radio galaxy in the 5.5 GHz and 9.0 GHz images. The red and blue dashed line represents the best-fit power-law model of J191108-595556 ($\alpha=-0.94\pm0.02$) and J191015-595615 ($\alpha=-0.81\pm0.03$) at 72-843 MHz.   }
    \label{fig:radiospec}
\end{figure}

The multiband counterparts of the AGN candidates in CXO1911 and CXO1910 are summarized in Table-\ref{tab:counterparts}. We calculated their fluxes (or magnitudes) and corresponding coordinates in X-ray, optical, and radio bands separately.  
To study the activities of the AGN candidates, we also illustrated the integrated radio spectrum of the sources in Figure-\ref{fig:radiospec}. J191108-595556 and J191015-595615 are found to have a spectral index of $\alpha=-1.14$ and $\alpha=-1.13$ at 72-231 MHz\citep{hurley2017}, which are roughly in agreement with the flux values measured at 843 MHz. At 5.5 GHz and 9.0 GHz, J191108-595556 is found to exhibit an excess of flux than the best-fit power-law model. The spectrum of CXOU J191100.49-595621.5 is found to be flat at 843 MHz, 5.5 GHz, and 9.0 GHz, suggesting that this source is young and the detected radio emission is dominated by the compact core of the radio galaxy. While for CXOU J191105.91-595558.8 and CXOU J191015.43-595624.9 (J191015-595615), their fluxes are found to decrease greatly at 9.0 GHz, indicating an older age for the plasma in radio jets. This picture is also consistent with the extended radio morphology observed in these two sources.

\section{Discussions and Conclusions}

In this paper, we report on the discovery of CXO1911 and CXO1910, two galaxy clusters serendipitously detected as extended X-ray sources with high significance levels ($S/N \sim 43$ and $S/N \sim 20$) in deep {\it Chandra} observations targeted to study globular cluster NGC 6752. Thanks to the deep ACIS-S observations with a total exposure time of $\sim 344$ ks, we detected at $3\sigma$ confidence level the rest frame 6.7-6.9 keV Iron {$K_{\alpha}$} line complex in the spectra of the clusters, from which we determined a redshift of $z=0.239\pm0.013$ for CXO1911 and $z=0.375\pm0.016$ for CXO1910. 
From spectral analysis, we also measured a global gas temperature of $kT=3.32^{+0.57}_{-0.46}$ keV, Fe abundance of $Z_{\rm Fe}=0.64^{+0.34}_{-0.29}Z_{\rm Fe\odot}$, and a rest-frame flux of $S_{0.5-7.0 \, {\rm keV}}=5.36^{+0.72}_{-0.62}\times 10^{-13} {\rm \, erg\, s^{-1}\, cm^{-2}}$ in CXO1911. While for CXO1910, these parameters are determined to be $kT=3.71^{+1.18}_{-0.86}$ keV, $Z_{\rm Fe}=1.29^{+0.97}_{-0.65}Z_{\rm Fe\odot}$ and $S_{0.5-7.0 \, {\rm keV}}=2.04^{+0.55}_{-0.45}\times 10^{-13} {\rm \, erg\, s^{-1}\, cm^{-2}}$, respectively.

The high-resolution {\it Chandra} data enabled us to perform spatially resolved analysis of the ICM in the galaxy clusters. We find significant X-ray cavities in the surface brightness images of CXO1911, which is consistent with the detection of cool core in the temperature profile and suggesting an effective AGN feedback in this cluster. The X-ray surface brightness profile of the galaxy clusters is well described by the $\beta$ model, from which we derived the deprojected electron density profile of the systems. We obtained an ICM mass of $M_{\rm ICM}(r<378\, {\rm kpc})=(4.7\pm1.0)\times 10^{12}\, M_{\odot}$ in CXO1911 and $M_{\rm ICM}(r<361\, {\rm kpc})=(4.2\pm1.7)\times 10^{12}\, M_{\odot}$ in CXO1910. Under the assumption of hydrostatic equilibrium, the total gravitating mass of the cluster is estimated to be $M(r<378\, {\rm kpc})=(7.9\pm2.4)\times 10^{13}\, M_{\odot}$ and $M(r<361\, {\rm kpc})=(9.2\pm5.8)\times 10^{13}\, M_{\odot}$ separately. 
Extrapolating the profile at larger radii, we find $M_{500}=(1.3\pm0.4)\times 10^{14}\, M_{\odot}$ for $r_{500}=705\pm72$ kpc in CXO1911 and $M_{500}=(2.0\pm1.5)\times 10^{14}\, M_{\odot}$ for $r_{500}=780\pm195$ kpc in CXO1910.

We search for the radio and optical counterparts of X-ray point sources detected within the FoVs of CXO1911 and CXO1910. Three sources, including CXOU J191100.49-595621.5 and CXOU J191105.91-595558.8 in CXO1911, and CXOU J191015.43-595624.9 in CXO1910, are found to have a counterpart radio source. CXOU J191105.91-595558.8 is further identified to be a narrow-angle-tail radio galaxy, with a pair of symmetric and elongated radio tails in the high-resolution ATCA images. CXOU J191100.49-595621.5 is found to be associated with an AGN, and its host galaxy is identified to be the central BCG of CXO1911. Although we have detected significant X-ray cavities around CXOU J191100.49-595621.5, the corresponding radio lobes are absent from the ATCA 5.5 GHz and 9.0 GHz images. These features may suggest an earlier episode of AGN feedback in this galaxy cluster. 

The absorbed flux of CXO1911 and CXO1910 in 0.5-2.0 keV band is $S_{0.5-2.0 \, {\rm keV}}=1.17^{+0.41}_{-0.42}\times 10^{-13} {\rm \, erg\, s^{-1}\, cm^{-2}}$ and $S_{0.5-2.0 \, {\rm keV}}=4.65^{+0.38}_{-0.36}\times 10^{-14} {\rm \, erg\, s^{-1}\, cm^{-2}}$, respectively. They are slightly larger than the detection threshold ($\sim 10^{-14} {\rm \, erg\, s^{-1}\, cm^{-2}}$, \citealp{Merloni2012, Liu2021}) of eROSITA, suggesting that both two systems could be detected (although CXO1910 may not be resolved as extended source) by eROSITA after its 4-years all-sky surveys. Considering that the FoV area of {\it Chandra} ACIS-S observations is $\sim 0.5\, {\rm deg^{2}}$, and the full sky area is $\sim 4.12 \times 10^{4}\, {\rm deg^{2}}$, the serendipitous detection of CXO1911 and CXO1910 supports that a large number ($\sim 10^{5}$) of X-ray luminous galaxy clusters will be detected by eROSITA telescope in the coming years \citep{Pillepich2012, Clerc2018}. 

\section*{Acknowledgements}
This work is supported by the Youth Program of National Natural Science Foundation of China No. 12003017, the Fundamental Research Funds for the Central Universities No. 2042020kf0035, the Natural Science Foundation of China under grants U1838103, No. 11622326 and No. 11763008, the National Program on Key Research and Development Project Grants No. 2016YFA0400803. Z.L. acknowledges support by the Fundamental Research Funds for the Central Universities of China.

\section*{Data Availability}

This research has used X-ray data obtained from the {\it Chandra} Data Archive\footnote{https://cda.harvard.edu/chaser} and software provided by the {\it Chandra} X-ray Center (CXC) in the application packages CIAO\footnote{http://cxc.harvard.edu/ciao}; the optical data products from the {\it HST} Data Archive\footnote{https://mast.stsci.edu/portal/Mashup/Clients/Mast/Portal.html}; and the ATCA data products from the Australia Telescope Online Archive (ATOA)\footnote{https://atoa.atnf.csiro.au/}.




\begin{thebibliography}{99}
\bibitem[\protect\citeauthoryear{Adami et al.}{2018}]{Adami2018} Adami C., Giles P., Koulouridis E., Pacaud F., Caretta C.~A., Pierre M., Eckert D., et al., 2018, A\&A, 620, A5. doi:10.1051/0004-6361/201731606
\bibitem[\protect\citeauthoryear{Allen, Evrard, \& Mantz}{2011}]{Allen2011} Allen S.~W., Evrard A.~E., Mantz A.~B., 2011, ARA\&A, 49, 409. doi:10.1146/annurev-astro-081710-102514
\bibitem[\protect\citeauthoryear{Arnaud}{1996}]{arnaud1996} Arnaud K.~A., 1996, ASPC, 101, 17
\bibitem[\protect\citeauthoryear{Arnaud et al.}{2010}]{arnaud2010} Arnaud M., Pratt G.~W., Piffaretti R., B{\"o}hringer H., Croston J.~H., Pointecouteau E., 2010, A\&A, 517, A92. doi:10.1051/0004-6361/200913416
\bibitem[\protect\citeauthoryear{Asplund et al.}{2009}]{asplund2009} Asplund M., Grevesse N., Sauval A.~J., Scott P., 2009, ARA\&A, 47, 481. doi:10.1146/annurev.astro.46.060407.145222
\bibitem[\protect\citeauthoryear{Begelman, Rees, \& Blandford}{1979}]{begelman1979} Begelman M.~C., Rees M.~J., Blandford R.~D., 1979, Natur, 279, 770. doi:10.1038/279770a0

\bibitem[\protect\citeauthoryear{B{\^\i}rzan et al.}{2004}]{birzan2004} B{\^\i}rzan L., Rafferty D.~A., McNamara B.~R., Wise M.~W., Nulsen P.~E.~J., 2004, ApJ, 607, 800. doi:10.1086/383519
\bibitem[\protect\citeauthoryear{B{\^\i}rzan et al.}{2012}]{birzan2012} B{\^\i}rzan L., Rafferty D.~A., Nulsen P.~E.~J., McNamara B.~R., R{\"o}ttgering H.~J.~A., Wise M.~W., Mittal R., 2012, MNRAS, 427, 3468. doi:10.1111/j.1365-2966.2012.22083.x
\bibitem[\protect\citeauthoryear{B{\^\i}rzan et al.}{2020}]{birzan2020} B{\^\i}rzan L., Rafferty D.~A., Br{\"u}ggen M., Botteon A., Brunetti G., Cuciti V., Edge A.~C., et al., 2020, MNRAS, 496, 2613. doi:10.1093/mnras/staa1594

\bibitem[\protect\citeauthoryear{Blanton et al.}{2009}]{blanton2009} Blanton E.~L., Randall S.~W., Douglass E.~M., Sarazin C.~L., Clarke T.~E., McNamara B.~R., 2009, ApJL, 697, L95. doi:10.1088/0004-637X/697/2/L95
\bibitem[\protect\citeauthoryear{Boehringer et al.}{1993}]{boehringer1993} Boehringer H., Voges W., Fabian A.~C., Edge A.~C., Neumann D.~M., 1993, MNRAS, 264, L25. doi:10.1093/mnras/264.1.L25
\bibitem[\protect\citeauthoryear{B{\"o}hringer et al.}{2000}]{bohringer2000} B{\"o}hringer H., Voges W., Huchra J.~P., McLean B., Giacconi R., Rosati P., Burg R., et al., 2000, ApJS, 129, 435. doi:10.1086/313427
\bibitem[\protect\citeauthoryear{B{\"o}hringer \& Werner}{2010}]{bohringer2010} B{\"o}hringer H., Werner N., 2010, A\&ARv, 18, 127. doi:10.1007/s00159-009-0023-3

\bibitem[\protect\citeauthoryear{Cash}{1979}]{cash1979} Cash W., 1979, ApJ, 228, 939. doi:10.1086/156922
\bibitem[\protect\citeauthoryear{Cavaliere \& Fusco-Femiano}{1976}]{cavaliere1976} Cavaliere A., Fusco-Femiano R., 1976, A\&A, 500, 95
\bibitem[\protect\citeauthoryear{Cheng et al.}{2018}]{cheng2018} Cheng Z., Li Z., Xu X., Li X., 2018, ApJ, 858, 33. doi:10.3847/1538-4357/aaba16
\bibitem[\protect\citeauthoryear{Cheng et al.}{2019}]{cheng2019} Cheng Z., Li Z., Li X., Xu X., Fang T., 2019, ApJ, 876, 59. doi:10.3847/1538-4357/ab1593
\bibitem[\protect\citeauthoryear{Clerc et al.}{2018}]{Clerc2018} Clerc N., Ramos-Ceja M.~E., Ridl J., Lamer G., Brunner H., Hofmann F., Comparat J., et al., 2018, A\&A, 617, A92. doi:10.1051/0004-6361/201732119
\bibitem[\protect\citeauthoryear{De Grandi \& Molendi}{2001}]{degrandi2001} De Grandi S., Molendi S., 2001, ApJ, 551, 153. doi:10.1086/320098
\bibitem[\protect\citeauthoryear{Dressler}{1980}]{dressler1980} Dressler A., 1980, ApJ, 236, 351. doi:10.1086/157753
\bibitem[\protect\citeauthoryear{Dressler et al.}{1997}]{dressler1997} Dressler A., Oemler A., Couch W.~J., Smail I., Ellis R.~S., Barger A., Butcher H., et al., 1997, ApJ, 490, 577. doi:10.1086/304890
\bibitem[\protect\citeauthoryear{Fabian}{1994}]{Fabian1994} Fabian A.~C., 1994, ARA\&A, 32, 277. doi:10.1146/annurev.aa.32.090194.001425
\bibitem[\protect\citeauthoryear{Fabian et al.}{2002}]{fabian2002} Fabian A.~C., Celotti A., Blundell K.~M., Kassim N.~E., Perley R.~A., 2002, MNRAS, 331, 369. doi:10.1046/j.1365-8711.2002.05182.x
\bibitem[\protect\citeauthoryear{Fabian et al.}{2006}]{fabian2006} Fabian A.~C., Sanders J.~S., Taylor G.~B., Allen S.~W., Crawford C.~S., Johnstone R.~M., Iwasawa K., 2006, MNRAS, 366, 417. doi:10.1111/j.1365-2966.2005.09896.x
\bibitem[\protect\citeauthoryear{Fabian}{2012}]{fabian2012} Fabian A.~C., 2012, ARA\&A, 50, 455. doi:10.1146/annurev-astro-081811-125521
\bibitem[\protect\citeauthoryear{Garon et al.}{2019}]{garon2019} Garon A.~F., Rudnick L., Wong O.~I., Jones T.~W., Kim J.-A., Andernach H., Shabala S.~S., et al., 2019, AJ, 157, 126. doi:10.3847/1538-3881/aaff62
\bibitem[\protect\citeauthoryear{Giacintucci \& Venturi}{2009}]{Giacintucci2009} Giacintucci S., Venturi T., 2009, A\&A, 505, 55. doi:10.1051/0004-6361/200912609
\bibitem[\protect\citeauthoryear{Hlavacek-Larrondo et al.}{2012}]{hlavacek2012} Hlavacek-Larrondo J., Fabian A.~C., Edge A.~C., Ebeling H., Sanders J.~S., Hogan M.~T., Taylor G.~B., 2012, MNRAS, 421, 1360. doi:10.1111/j.1365-2966.2011.20405.x
\bibitem[\protect\citeauthoryear{Hlavacek-Larrondo et al.}{2015}]{hlavacek2015} Hlavacek-Larrondo J., McDonald M., Benson B.~A., Forman W.~R., Allen S.~W., Bleem L.~E., Ashby M.~L.~N., et al., 2015, ApJ, 805, 35. doi:10.1088/0004-637X/805/1/35
\bibitem[\protect\citeauthoryear{Hurley-Walker et al.}{2017}]{hurley2017} Hurley-Walker N., Callingham J.~R., Hancock P.~J., Franzen T.~M.~O., Hindson L., Kapi{\'n}ska A.~D., Morgan J., et al., 2017, MNRAS, 464, 1146. doi:10.1093/mnras/stw2337
\bibitem[\protect\citeauthoryear{Jones \& Owen}{1979}]{jones1979} Jones T.~W., Owen F.~N., 1979, ApJ, 234, 818. doi:10.1086/157561
\bibitem[\protect\citeauthoryear{Koulouridis et al.}{2021}]{Koulouridis2021} Koulouridis E., Clerc N., Sadibekova T., Chira M., Drigga E., Faccioli L., Le F{\`e}vre J.~P., et al., 2021, arXiv, arXiv:2104.06617
\bibitem[\protect\citeauthoryear{Leccardi \& Molendi}{2008}]{leccardi2008} Leccardi A., Molendi S., 2008, A\&A, 486, 359. doi:10.1051/0004-6361:200809538

\bibitem[\protect\citeauthoryear{Liedahl, Osterheld, \& Goldstein}{1995}]{liedahl1995} Liedahl D.~A., Osterheld A.~L., Goldstein W.~H., 1995, ApJL, 438, L115. doi:10.1086/187729
\bibitem[\protect\citeauthoryear{Li et al.}{2011}]{Li2011} Li Z., Jones C., Forman W.~R., Kraft R.~P., Lal D.~V., Di Stefano R., Spitler L.~R., et al., 2011, ApJ, 730, 84. doi:10.1088/0004-637X/730/2/84
\bibitem[\protect\citeauthoryear{Liu et al.}{2021}]{Liu2021} Liu A., Bulbul E., Ghirardini V., Liu T., Klein M., Clerc N., Oezsoy Y., et al., 2021, arXiv, arXiv:2106.14518
\bibitem[\protect\citeauthoryear{Kaastra}{1992}]{kaastra1992} Kaastra, J. 1992, An X-Ray Spectral Code for Optically Thin Plasmas (Internal SRON-Leiden Report, updated version 2.0)
\bibitem[\protect\citeauthoryear{Kravtsov \& Borgani}{2012}]{Kravtsov2012} Kravtsov A.~V., Borgani S., 2012, ARA\&A, 50, 353. doi:10.1146/annurev-astro-081811-125502

\bibitem[\protect\citeauthoryear{Machacek et al.}{2011}]{machacek2011} Machacek M.~E., Jerius D., Kraft R., Forman W.~R., Jones C., Randall S., Giacintucci S., et al., 2011, ApJ, 743, 15. doi:10.1088/0004-637X/743/1/15
\bibitem[\protect\citeauthoryear{Mao et al.}{2009}]{mao2009} Mao M.~Y., Johnston-Hollitt M., Stevens J.~B., Wotherspoon S.~J., 2009, MNRAS, 392, 1070. doi:10.1111/j.1365-2966.2008.14141.x

\bibitem[\protect\citeauthoryear{Mauch et al.}{2003}]{mauch2003} Mauch T., Murphy T., Buttery H.~J., Curran J., Hunstead R.~W., Piestrzynski B., Robertson J.~G., et al., 2003, MNRAS, 342, 1117. doi:10.1046/j.1365-8711.2003.06605.x
\bibitem[\protect\citeauthoryear{McNamara et al.}{2000}]{mcnamara2000} McNamara B.~R., Wise M., Nulsen P.~E.~J., David L.~P., Sarazin C.~L., Bautz M., Markevitch M., et al., 2000, ApJL, 534, L135. doi:10.1086/312662

\bibitem[\protect\citeauthoryear{McNamara \& Nulsen}{2007}]{mcnamara2007} McNamara B.~R., Nulsen P.~E.~J., 2007, ARA\&A, 45, 117. doi:10.1146/annurev.astro.45.051806.110625
\bibitem[\protect\citeauthoryear{McNamara \& Nulsen}{2012}]{mcnamara2012} McNamara B.~R., Nulsen P.~E.~J., 2012, NJPh, 14, 055023. doi:10.1088/1367-2630/14/5/055023
\bibitem[\protect\citeauthoryear{Merloni et al.}{2012}]{Merloni2012} Merloni A., Predehl P., Becker W., B{\"o}hringer H., Boller T., Brunner H., Brusa M., et al., 2012, arXiv, arXiv:1209.3114
\bibitem[\protect\citeauthoryear{Nousek \& Shue}{1989}]{nousek1989} Nousek J.~A., Shue D.~R., 1989, ApJ, 342, 1207. doi:10.1086/167676
\bibitem[\protect\citeauthoryear{Pierre et al.}{2016}]{Pierre2016} Pierre M., Pacaud F., Adami C., Alis S., Altieri B., Baran N., Benoist C., et al., 2016, A\&A, 592, A1. doi:10.1051/0004-6361/201526766
\bibitem[\protect\citeauthoryear{Pillepich, Porciani, \& Reiprich}{2012}]{Pillepich2012} Pillepich A., Porciani C., Reiprich T.~H., 2012, MNRAS, 422, 44. doi:10.1111/j.1365-2966.2012.20443.x
\bibitem[\protect\citeauthoryear{Planelles et al.}{2014}]{planelles2014} Planelles S., Borgani S., Fabjan D., Killedar M., Murante G., Granato G.~L., Ragone-Figueroa C., et al., 2014, MNRAS, 438, 195. doi:10.1093/mnras/stt2141

\bibitem[\protect\citeauthoryear{Poggianti et al.}{1999}]{poggianti1999} Poggianti B.~M., Smail I., Dressler A., Couch W.~J., Barger A.~J., Butcher H., Ellis R.~S., et al., 1999, ApJ, 518, 576. doi:10.1086/307322
\bibitem[\protect\citeauthoryear{Pratt et al.}{2007}]{pratt2007} Pratt G.~W., B{\"o}hringer H., Croston J.~H., Arnaud M., Borgani S., Finoguenov A., Temple R.~F., 2007, A\&A, 461, 71. doi:10.1051/0004-6361:20065676
\bibitem[\protect\citeauthoryear{Rafferty, McNamara, \& Nulsen}{2008}]{rafferty2008} Rafferty D.~A., McNamara B.~R., Nulsen P.~E.~J., 2008, ApJ, 687, 899. doi:10.1086/591240
\bibitem[\protect\citeauthoryear{Randall et al.}{2011}]{randall2011} Randall S.~W., Forman W.~R., Giacintucci S., Nulsen P.~E.~J., Sun M., Jones C., Churazov E., et al., 2011, ApJ, 726, 86. doi:10.1088/0004-637X/726/2/86
\bibitem[\protect\citeauthoryear{Sanders, Fabian, \& Taylor}{2009}]{sanders2009} Sanders J.~S., Fabian A.~C., Taylor G.~B., 2009, MNRAS, 393, 71. doi:10.1111/j.1365-2966.2008.14207.x
\bibitem[\protect\citeauthoryear{Sarazin}{1986}]{sarazin1986} Sarazin C.~L., 1986, RvMP, 58, 1. doi:10.1103/RevModPhys.58.1
\bibitem[\protect\citeauthoryear{Sarazin}{1988}]{sarazin1988} Sarazin C.~L., 1988, xrec.book
\bibitem[\protect\citeauthoryear{Shin, Woo, \& Mulchaey}{2016}]{shin2016} Shin J., Woo J.-H., Mulchaey J.~S., 2016, ApJS, 227, 31. doi:10.3847/1538-4365/227/2/31
\bibitem[\protect\citeauthoryear{Sun et al.}{2009}]{sun2009} Sun M., Voit G.~M., Donahue M., Jones C., Forman W., Vikhlinin A., 2009, ApJ, 693, 1142. doi:10.1088/0004-637X/693/2/1142
\bibitem[\protect\citeauthoryear{Vikhlinin et al.}{1998}]{Vikhlinin1998} Vikhlinin A., McNamara B.~R., Forman W., Jones C., Quintana H., Hornstrup A., 1998, ApJ, 502, 558. doi:10.1086/305951

\bibitem[\protect\citeauthoryear{Vikhlinin et al.}{2005}]{Vikhlinin2005} Vikhlinin A., Markevitch M., Murray S.~S., Jones C., Forman W., Van Speybroeck L., 2005, ApJ, 628, 655. doi:10.1086/431142
\bibitem[\protect\citeauthoryear{Vikhlinin et al.}{2006}]{Vikhlinin2006} Vikhlinin A., Kravtsov A., Forman W., Jones C., Markevitch M., Murray S.~S., Van Speybroeck L., 2006, ApJ, 640, 691. doi:10.1086/500288

\bibitem[\protect\citeauthoryear{Vikhlinin et al.}{2009}]{Vikhlinin2009} Vikhlinin A., Burenin R.~A., Ebeling H., Forman W.~R., Hornstrup A., Jones C., Kravtsov A.~V., et al., 2009, ApJ, 692, 1033. doi:10.1088/0004-637X/692/2/1033
\bibitem[\protect\citeauthoryear{Wellington, Miley, \& van der Laan}{1973}]{Wellington1973} Wellington K.~J., Miley G.~K., van der Laan H., 1973, Natur, 244, 502. doi:10.1038/244502a0
\bibitem[\protect\citeauthoryear{Willingale et al.}{2013}]{Willingale2013} Willingale R., Starling R.~L.~C., Beardmore A.~P., Tanvir N.~R., O'Brien P.~T., 2013, MNRAS, 431, 394. doi:10.1093/mnras/stt175
\bibitem[\protect\citeauthoryear{Wing \& Blanton}{2011}]{Wing2011} Wing J.~D., Blanton E.~L., 2011, AJ, 141, 88. doi:10.1088/0004-6256/141/3/88

\bibitem[\protect\citeauthoryear{Yu et al.}{2011}]{yu2011} Yu H., Tozzi P., Borgani S., Rosati P., Zhu Z.-H., 2011, A\&A, 529, A65. doi:10.1051/0004-6361/201016236


\end{thebibliography}








\bsp	
\label{lastpage}
\end{document}